\documentclass[%
 reprint,
superscriptaddress,
nofootinbib,
nobibnotes,
 amsmath,amssymb,
 aps,
prb,
]{revtex4-2}
\usepackage[utf8]{inputenc}
\usepackage{amsmath}
\usepackage{amsfonts}
\usepackage{amssymb}
\usepackage{graphicx}
\usepackage[caption=false]{subfig}
\usepackage{hyperref}
\usepackage{braket}
\usepackage[export]{adjustbox}
\usepackage[colorinlistoftodos]{todonotes}
\usepackage[left=2cm,right=2cm,top=2cm,bottom=2cm]{geometry}
\usepackage{soul}
\usepackage{float}
\newcommand{\bnu}{\nu_0}

\begin{document}

\title{The origins of noise in the Zeeman splitting of spin qubits in natural-silicon devices}
\author{J. S. Rojas-Arias}
\email{juan.rojasarias@riken.jp}
\affiliation{RIKEN, Center for Quantum Computing (RQC), Wako-shi, Saitama 351-0198, Japan}
\author{Y. Kojima}
\author{K. Takeda}
\affiliation{RIKEN, Center for Emergent Matter Science (CEMS), Wako-shi, Saitama 351-0198, Japan}
\author{P. Stano}
\affiliation{RIKEN, Center for Emergent Matter Science (CEMS), Wako-shi, Saitama 351-0198, Japan}
\affiliation{Slovak Academy of Sciences, Institute of Physics, 845 11 Bratislava, Slovakia}
\author{T. Nakajima}
\affiliation{RIKEN, Center for Emergent Matter Science (CEMS), Wako-shi, Saitama 351-0198, Japan}
\author{J. Yoneda}
\affiliation{Tokyo Institute of Technology, Tokyo Tech Academy for Super Smart Society, Tokyo 152-8552, Japan}
\author{A. Noiri}
\affiliation{RIKEN, Center for Emergent Matter Science (CEMS), Wako-shi, Saitama 351-0198, Japan}
\author{T. Kobayashi}
\affiliation{RIKEN, Center for Quantum Computing (RQC), Wako-shi, Saitama 351-0198, Japan}
\author{D. Loss}
\affiliation{RIKEN, Center for Quantum Computing (RQC), Wako-shi, Saitama 351-0198, Japan}
\affiliation{RIKEN, Center for Emergent Matter Science (CEMS), Wako-shi, Saitama 351-0198, Japan}
\affiliation{Department of Physics, University of Basel, Klingelbergstrasse 82, CH-4056 Basel, Switzerland}
\author{S. Tarucha}
\email{tarucha@riken.jp}
\affiliation{RIKEN, Center for Quantum Computing (RQC), Wako-shi, Saitama 351-0198, Japan}
\affiliation{RIKEN, Center for Emergent Matter Science (CEMS), Wako-shi, Saitama 351-0198, Japan}

\begin{abstract}

We measure and analyze noise-induced energy-fluctuations of spin qubits defined in quantum dots made of isotopically natural silicon. Combining Ramsey, time-correlation of single-shot measurements, and CPMG experiments, we cover the qubit noise power spectrum over a frequency range of nine orders of magnitude without any gaps. We find that the low-frequency noise spectrum is similar across three different devices suggesting that it is dominated by the hyperfine coupling to nuclei. The effects of charge noise are smaller, but not negligible, and are device dependent as confirmed from the noise cross-correlations. We also observe differences to spectra reported in GaAs {[Phys. Rev. Lett. 118, 177702 (2017), Phys. Rev. Lett. 101, 236803 (2008)]}, which we attribute to the presence of the valley degree of freedom in silicon. Finally, we observe $T_2^*$ to increase upon increasing the external magnetic field, which we speculate is due to the increasing field-gradient of the micromagnet suppressing nuclear spin diffusion.
\end{abstract}

\maketitle

\section{Introduction}

Among the candidates for the realization of a quantum computer, qubits defined in the spin states of electrons confined in quantum dots hold great promise due to their small size \cite{Loss1998,Burkard2023}. Initially fabricated mostly in GaAs \cite{Petta2005}, about a decade later the field has largely shifted towards Si as a better host for this type of qubits \cite{Maune2012,Prance2012}. Besides compatibility with technology for classical electronics, the main reason to use Si is the lower noise due to hyperfine coupling to nuclear spins in the material \cite{Zwanenburg2013}.

Natural silicon has a high concentration of spinless nuclei, with only $4.7$\% of $^{29}$Si, the only spinfull isotope. Consequently, qubits in Si have lower nuclear-spin noise compared to GaAs with no spinless isotopes. By changing the material from GaAs to Si, the coherence of spin qubits moved from $\sim10$ ns to $\sim1\ \mu$s \cite{Stano2022}. Moreover, silicon can be isotopically purified, decreasing the fraction of $^{29}$Si and increasing the coherence time to the order of tens or hundreds of $\mu$s \cite{Stano2022, Cvitkovich2024}. 

Seeing this relation between the spin-qubit coherence and nuclear spin concentration, it is reasonable to expect that natural silicon samples are limited primarily by nuclear noise.  However, confirming this expectation experimentally is not straightforward, as it is not easy to tell apart nuclear and charge noise. We have recently shown that the analysis of qubit noise cross-correlations might serve this purpose \cite{Yoneda2023}. Surprisingly, we found that in that experiment with a natural-silicon device, charge noise was dominant, in contradiction with the above given expectation. Here, we examine the question of dominant noise in natural silicon in detail, using multiple devices. 

Specifically, we analyze noise auto- and cross-spectra in devices with different metallic gate layouts and measurement setups. At low frequencies ($<10^2$ Hz), we find a device-independent (`universal') noise spectrum across the measured seven qubits in three devices. We believe that in this range nuclear noise dominates. However, the analysis of cross-spectra reveals a sizable contribution of device-dependent charge noise. At frequencies above 10 kHz, the noise auto-spectra are no longer universal and are likely dominated by charge noise.

An important part of our investigations is a novel spectroscopy method that we implement. It is based on the correlation of single-shot measurements and allows us to examine the noise in the otherwise inaccessible frequency range (roughly 10 Hz -- 10 kHz). Filling this gap is critical to cross-check that we obtain a consistent qubit noise spectrum in the range covering 9 orders of magnitude in frequency. 

The paper is organized as follows. In Sec.~\ref{sec:device} we introduce the devices. In Sec.~\ref{sec:universal} we present the qubit low-frequency noise auto-spectra and analyze their origin. In Sec.~\ref{sec:cross} we quantify the charge-noise contribution using the cross-spectra. In Sec.\ref{sec:wide} we describe and illustrate a novel noise spectroscopy for mid-range frequencies. In Sec.~\ref{sec:b_dependence} we present the dependence of the dephasing time $T_2^*$ on the magnetic field.

\section{Devices}\label{sec:device}

\begin{figure}
	\subfloat{\includegraphics[width=0.5\columnwidth,valign=c]{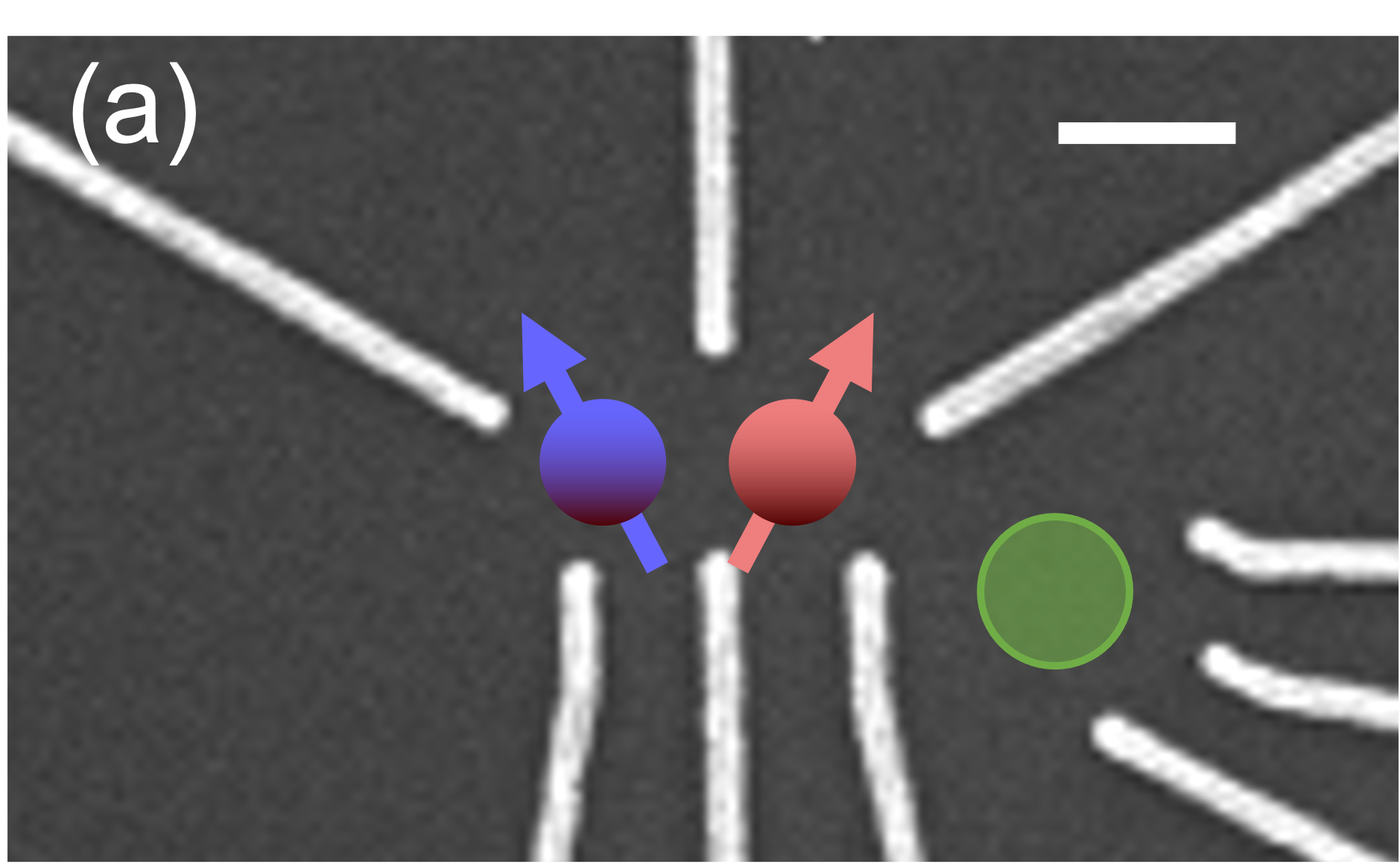}\label{fig:device_12}
		}
	\subfloat{\includegraphics[width=0.5\columnwidth,valign=c]{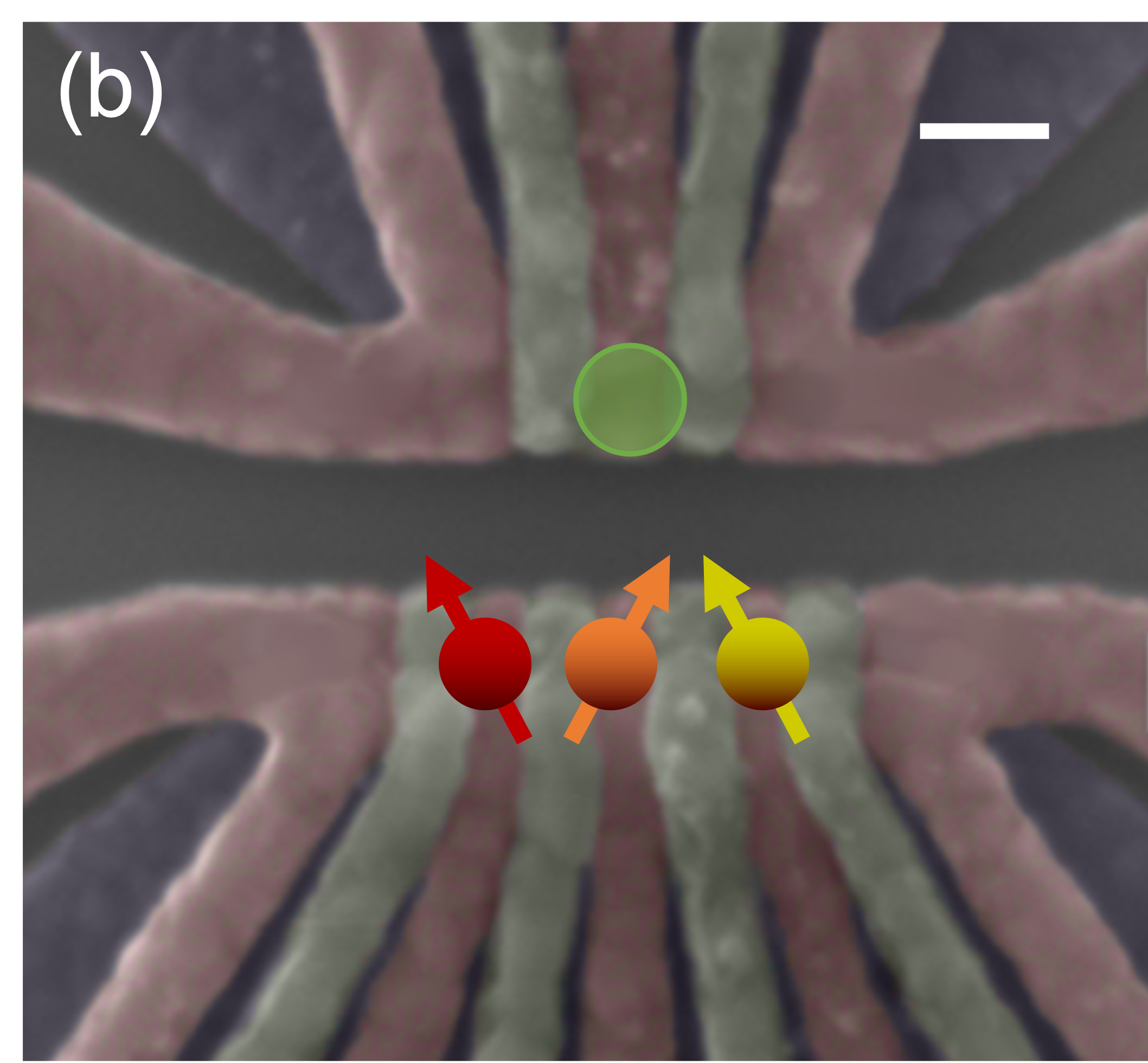}\
\label{fig:device_3}
		}
\caption{A false-color scanning electron microscope images of devices nominally identical to the ones measured, (a) devices D1 and D2, and (b) device D3. In (a) the structure is covered by a global top gate (not shown) for accumulation of charges. The white scale bars indicate 100 nm. Spin qubits are depicted as arrows with the green circle depicting the charge sensor.}
\label{fig:system}
\end{figure}

We measure three devices, labeled D1, D2, and D3. All are made from the same Si/SiGe heterostructure with a 15 nm wide Si quantum well topped by a 60 nm SiGe spacer. The heterostructure was made with natural silicon, meaning  4.7\% of $^{29}$Si isotopes with spin. Devices D1 and D2 host $N=2$ qubits, with the gate structure shown in Fig.~\ref{fig:device_12}. Device D3 uses overlapping gates with $N=3$ qubits \cite{Takeda2022} shown in Fig.~\ref{fig:device_3}. Qubits are labeled L, C, and R, referring to their location (left, center, and right) with qubit C only present in the three-qubit device. All devices contain a Co micromagnet on top, isolated from the metallic gates by an Al$_2$O$_3$ layer. It induces a magnetic field gradient serving as an artificial spin-orbit interaction to enable electrical manipulation of the spins. The micromagnet is polarized by an externally applied in-plane magnetic field perpendicular to the line defined by the qubits. 

The readout is done through Pauli spin blockade (PSB) for D1 and D2, and through energy selective tunneling for D3 as described in Ref.~\cite{Takeda2022}. The charge state of a dot array is detected by a sensor quantum dot (green circles in Fig.~\ref{fig:system}) connected to a tank circuit for radio-frequency reflectometry \cite{Noiri2020}.

\section{Universal noise power spectrum}\label{sec:universal}

To examine the low-frequency noise in the Zeeman splittings of the qubits, referred to hereafter as `qubit energies', we conducted an interleaved Ramsey experiment. A single-qubit Ramsey sequence consists of spin initialization, a $\pi/2$-rotation, a free evolution for time $\tau$, a second $\pi/2$-rotation, and spin readout. We apply identical (meaning a fixed $\tau$) single-qubit sequence to the $N$ qubits sequentially ($N=2$ for D1 and D2, $N=3$ for D3) and repeat it for increasing values of $\tau$ from $\tau_\mathrm{min}$ to $\tau_\mathrm{max}$ in steps $\Delta\tau$. Data collected from such a scan of $N$ (inner loop) and $\tau$ (outer loop) are called one `record'. Using Bayesian estimation \cite{Delbecq2016, Nakajima2020}, for each qubit in the array, we assign one energy value to each record, in turn to a (wall) time corresponding to that record acquisition. In this way, we obtain the (wall) time traces of qubit energies, that is their estimated values at discrete equidistant times. Device-specific details on these sequences are in Appendix~\ref{app:details}.

\begin{figure}
\includegraphics[width=\columnwidth]{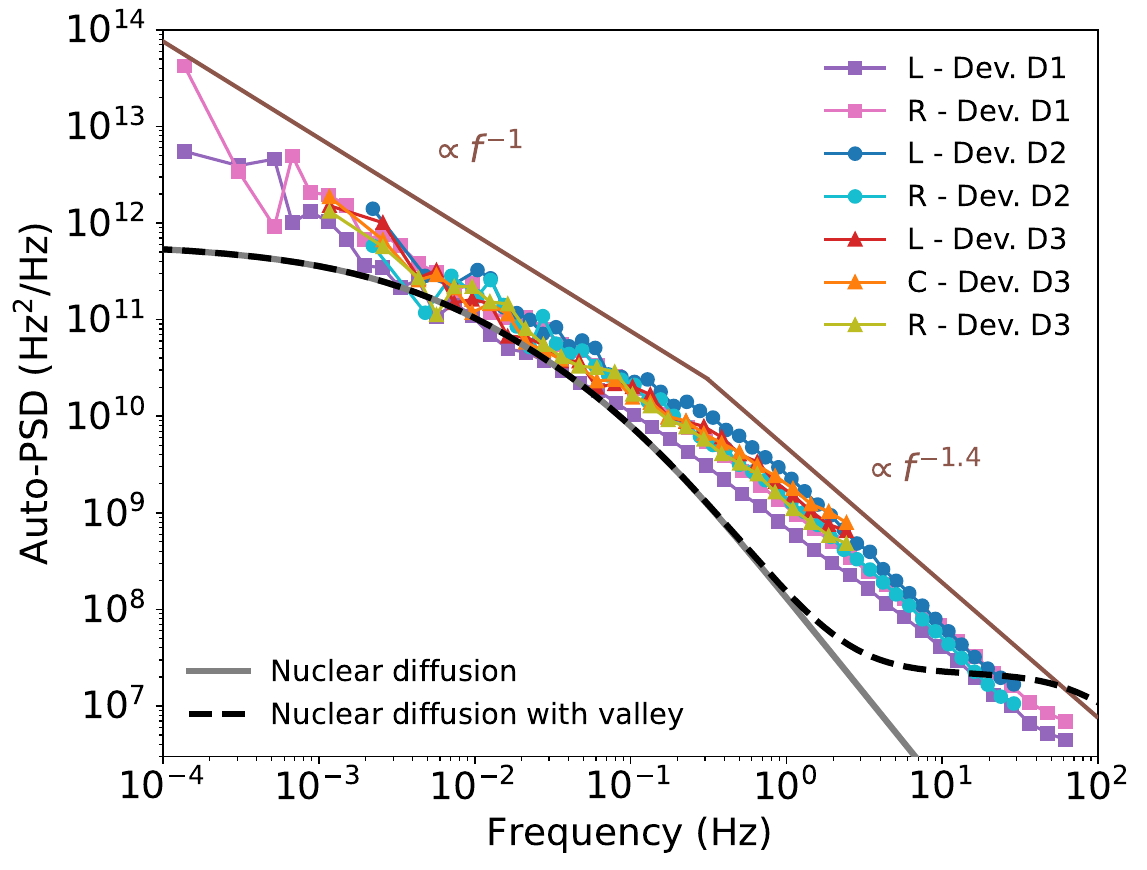}
\caption{Noise auto-correlation spectrum for qubit energies (seven qubits in three devices as labeled). The brown lines show $f^{-1}$ and $f^{-1.4}$ dependence for reference. Gray continuous and black dashed lines are the theory predictions based on the nuclear diffusion model without and with valley oscillations in the electron's wave function, respectively. The model assumes a Gaussian wavefunction with confinement lengths $l_p=13.9$ nm in the plane and $l_y=2.7$ nm out of the plane. The former corresponds to an in-plane confinement strength of 2 meV. The latter value is fixed by requiring the effective volume $(\int dV |\psi|^4)^{-1}$ to be the same as for the ground-state of a hard-wall confinement of 15 nm width, the actual width of the quantum well.}
\label{fig:nuclear_psd}
\end{figure}

From the time traces of the qubit energies, we evaluate the auto power spectral density (auto-PSD) using the method of Ref.~\cite{Gutierrez-Rubio2022}. The resulting auto-PSDs are displayed in Figure~\ref{fig:nuclear_psd}. Surprisingly, the spectra have essentially the same shape for all qubits, even across different devices. We will refer to this uniformity as the auto-PSD being ``universal''. The insensitivity to device details suggests that the dominant source of noise is the common element: The material itself and, specifically, its magnetic noise from $^{29}$Si nuclear spins. While we can not exclude the possibility that we observe some unidentified charge noise specific to the silicon wafer \footnote{We rule out noise intrinsic to the setup as device D3 was measured in a separate dilution fridge with different equipment.}, we have further reasons against it: In Appendix~\ref{app:sensor}, we show the auto-PSD of the charge sensor. In contrast to Figure~\ref{fig:nuclear_psd}, it roughly follows a $f^{-1}$ possibly crossing over to a white noise floor at the highest measured frequencies. Were wafer-originated charge noise the source of universal noise for qubits, it should also show up in the charge sensor spectrum. Since this is not the case, we conclude that charge noise is probably not the primary contributor to the observed noise spectra.

Figure~\ref{fig:nuclear_psd} displays a noise spectrum with a $f^{-1}$ trend at low frequencies changing to $f^{-1.4}$ around $0.4$ Hz. This behavior deviates from the theory based on the three-dimensional diffusion model. The latter predicts a curve that is a constant in the zero frequency limit, changing smoothly to a $f^{-2}$ fall-off in the high-frequency limit. The high-frequency fall-off was observed in GaAs quantum dots in Refs.~\cite{Malinowski2017} and \cite{Reilly2008}. Using the materials parameters for silicon, we plot the noise spectrum predicted by this model in gray in Fig.~\ref{fig:nuclear_psd}, evaluating the model given in Ref.~\cite{Reilly2008}. While the prediction has the correct order of magnitude, it differs from the observed spectra at low and high frequencies. In an attempt to improve the correspondence, we have revisited this standard diffusion model by including the silicon valley degree of freedom (see Appendix~\ref{app:nuclear}). Plotting it in black dashed in Fig.~\ref{fig:nuclear_psd},   we find that the valley has visible effects at higher frequencies, diminishing the theory-measurement discrepancy. We leave the remaining differences unexplained, speculating that they are due to the limits of the applicability of the diffusion model assumptions. In any case, reproducing the correct order of magnitude as well as the shape qualitatively gives further support to assigning the spectra in Fig.~\ref{fig:nuclear_psd} to nuclear noise.

\section{Charge noise contribution}\label{sec:cross}

We gave several arguments in favor of assigning the observed auto-PSDs to hyperfine noise: the similarity across qubits and devices, qualitative agreement with the diffusion model, and difference to the charge-sensor spectrum. In this section, we analyze cross-power spectral densities (cross-PSD) as an additional probe into the noise character. It allows us to access the charge-noise contribution by filtering out the hyperfine noise. This separation is possible because the hyperfine noise is local: coming from the contact hyperfine interaction, its range is set by the electron wavefunction. We have estimated the resulting non-local part in Refs.~\cite{Yoneda2023,Rojas-Arias2023} and found that, assuming the diffusion model, the hyperfine noise gives orders of magnitude smaller cross-PSD compared to the auto-PSD. In contrast, electrical noise is long range and induces sizable correlations, in magnitude comparable to the auto-PSD. Using cross-PSD, which we have access to as we track the qubits' energies simultaneously, we can thus isolate the charge-noise effects. 

\begin{figure}
\centering
\subfloat{\includegraphics[width=0.9\columnwidth]{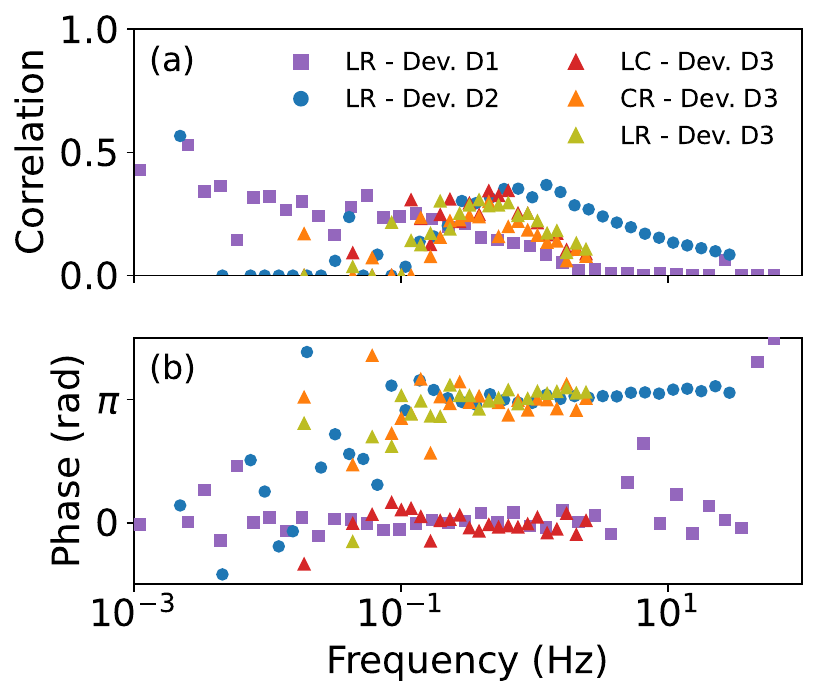}\label{fig:correlated_collection}}
\subfloat{\label{fig:correlated_collection_phase}}
\caption{Collection of cross-PSDs for all qubit pairs in each device. Normalized magnitude (a) and phase (b) of the correlations.}
\label{fig:cross_collection}
\end{figure}

In Fig.~\ref{fig:cross_collection} we present a collection of both the magnitude (normalized with respect to the corresponding auto-PSDs) and phase of the cross-PSDs between qubit pairs. The presence of non-negligible correlations proves that the spectra in Fig.~\ref{fig:nuclear_psd} have sizable contribution from charge noise. An important difference is that the cross-PSDs are device dependent. It suggests that 1) the auto- and cross-PSDs are dominated by different phenomena, and 2) the charge-noise sources are near the qubits. Assigning the charge noise to a few local sources was the conclusion of Refs.~\cite{Rojas-Arias2023,Connors2022,Gungordu2019} and we believe that the same applies here. 

We note that in all of our devices, as well as in Refs.~\cite{Yoneda2023} and \cite{Rojas-Arias2023}, the phase of the cross-PSD is close to 0 or $\pi$ at the points where the correlation amplitude is sizable. Instances where the phase appears to take other values are those where the correlation amplitude is close to 0, where, naturally, the phase randomizes and is irrelevant. Thus, we concude that the cross-PSD takes on real values, meaning that the source of correlated noise affects both qubits simultaneously without any delay (on the scale given by the inverse of the corresponding frequency).

For all pairs, the maximal correlation is at least $0.4$, though reached at different frequencies. Since not all charge noise is necessarily correlated, this value is a lower bound. However, because the variation in the cross-PSDs across devices is not seen in the auto-PSDs, we believe that the actual value is not much larger than the lower bound. We conclude that while most of the noise is from nuclear spins, in every device there are frequency ranges where  $\sim40$\% of the low-frequency noise is due to charge noise. We note that device-specific factors and tuning can influence this balance, as shown in Ref.~\cite{Yoneda2023} where charge noise was dominant for one of the qubits, while a combination of both noise sources was observed in the other.

Before concluding this section, we look at the dependence of the correlations on the dot-dot distance using the data from device D3, which contains a non-neighboring qubit pair L-R. Interestingly, correlations do not appreciably decay with distance within the range we can access: The correlations between nearest neighbors (L-C and C-R) and next-to-nearest neighbors (L-R) are similar. The correlation phases are consistent in the sense that if L-C are positively correlated and C-R are anticorrelated, L-R should be anticorrelated, which is the case. However, these correlations are not due to an electric field constant across the device (and fluctuating in time). Indeed, given that all qubits are subject to the same magnetic gradient, a global fluctuating electric field would affect all qubits similarly leading to all cross correlations being positive, in contradiction to our observations. Instead, it is likely that these correlations arise from a charge two-level system (TLS) located close to the qubits. We draw further support for this guess from observing a Lorentzian shape typical for a TLS in the unnormalized cross-PSD magnitude (see Appendix~\ref{app:lorentzian}). With this interpretation, we expect that correlations would decay on the scale of the mean distance between TLSs \cite{Rojas-Arias2023}. The L-R qubits' distance of ~200 nm gives $n_c\sim10^9$ cm${}^{-2}$ as a ballpark estimate of the TLS density. Given the impact of long-ranged noise correlations on quantum error correction, we close by stressing that more studies are needed on the scaling of noise correlations in larger qubit arrays.

\section{Noise spectroscopy from correlation of single-shot measurements}\label{sec:wide}

\subsection{A gap in the noise spectrum}

The frequency range at which the qubit noise can be obtained from Ramsey experiments is limited. While it is conceptually straightforward to explore lower frequencies, by extending the total measurement time, very long times are both impractical and might be beyond the stability of the device. Probing higher frequencies is even more difficult. It requires to reduce the total measurement time of a record. That can be done either by shortening the readout time (the longest part of an individual cycle), which worsens measurement fidelity, or decreasing the number of measurements within the record, which worsens the estimation fidelity. We have found that 10 - 100 Hz is the upper frequency limit achievable by the Ramsey experiment. To extract the noise at higher frequencies, we resort to other techniques.

\begin{figure}
\centering
\includegraphics[width=\columnwidth]{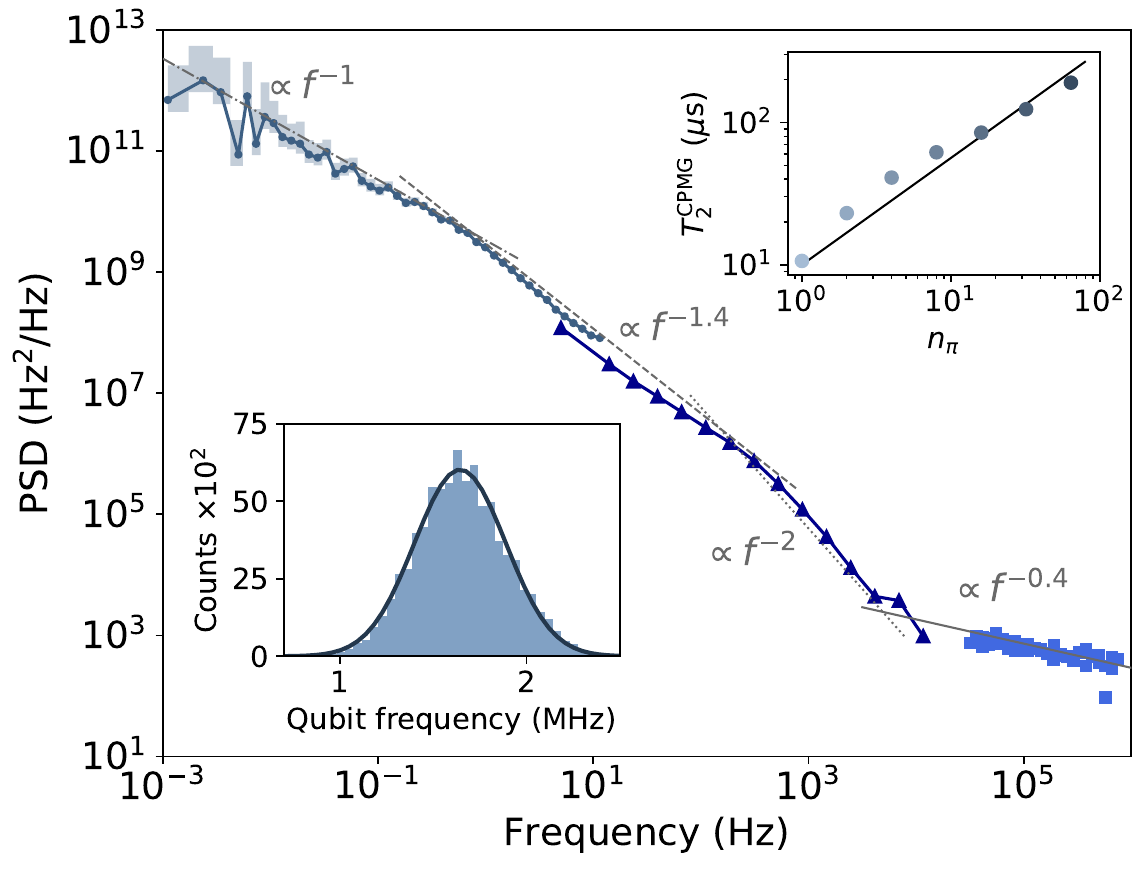}
\caption{Noise auto-PSD of qubit L of device D2. The spectrum is obtained by combining three different methods: correlation of qubit energies via Bayesian estimation (connected dots), time correlation of single-shot measurements (connected triangles), and CPMG dynamical decoupling (squares). For the Bayesian method, confidence intervals of 90\% are presented as shaded areas. Continuous, dotted, dashed, and dashed-dotted lines show $f^{-0.4}$, $f^{-2}$, $f^{-1.4}$, and $f^{-1}$ dependence for reference, respectively. Upper right inset: coherence time for different number of pulses of the CPMG sequence. Lower left inset: histogram of qubit energies with a Gaussian fit as a continuous curve.}
\label{fig:wide}
\end{figure}

First, we implement dynamical decoupling as an established method to probe noise at around MHz frequencies \cite{Yoneda2018,Kawakami2016}. We use the Carr-Purcell-Meiboom-Gill (CPMG) protocol, applying between 1 to 64 of $\pi$-rotations during the evolution time \cite{Cywinski2008} for qubit $L$ in device D2. The inset in Fig.~\ref{fig:wide} illustrates the resulting increase in qubit coherence, from the inhomogeneous dephasing time (called `coherence time' in further, for conciseness) $T_2^*=1.03\ \mu$s up to $T_2^\mathrm{CPGM}=190\ \mu$s. The large increase confirms the dominance of low-frequency noise seen in Fig.~\ref{fig:nuclear_psd}. More importantly here, the CPMG decay curves (not shown) allow us to extract the auto-PSD \cite{Yoneda2018,Connors2022}. We are able to resolve the auto-PSD in the frequency range $10^{4}\ \mathrm{Hz}\leq f\leq10^{6}\ \mathrm{Hz}$, the result is displayed as squares in Fig.~\ref{fig:wide}. We observe a spectrum decaying as $f^{-0.4}$, much flatter than the low-frequency part.

Extrapolating the $f^{-1.4}$ low-frequency and $f^{-0.4}$ high-frequency trends into the mid-frequency range does not lead to a consistent picture and casts doubts on the two procedures reliability. It strongly motivates finding a way to access the noise at intermediate frequencies. However, similarly as extending the Ramsey experiment to higher frequencies, extending the CPMG to lower frequencies is not feasible. The lowest achievable frequency is limited by the lack of resolution of the CPMG decay curves at evolution times longer than the coherence time.

We are thus left with a gap for frequencies between $\sim10$ Hz and $\sim10^4$ Hz. Such a gap in the noise spectrum at intermediate frequencies can be found in previously reported noise spectra \cite{Yoneda2018,Nakajima2020,Jock2022,Camenzind2022}. In Ref.~\cite{Connors2022}, this gap was filled by presenting the noise spectrum of the charge sensor. However, this solution relies on the assumption that qubit noise is identical to the charge sensor signal noise and, therefore, must be dominated by charge noise in the first place. The assumption is difficult to prove if fulfilled, but easy to disprove if not. The latter is the case here and the solution from Ref.~\cite{Connors2022} is not available. 

Instead,  we implement a method that directly probes the qubit noise at intermediate frequencies. We build on the ideas from Ref.~\cite{Fink2013}, which proposed extracting the noise spectrum from time correlations of single-shot measurements. Trying to apply the proposal of Ref.~\cite{Fink2013}, we found that the method needs adjustments for measurement errors and other minor refinements. With these, we have succeeded in implementing the proposal---as far as we know---for the first time on real experimental data. 

\subsection{Filling the gap with single-shot measurements}

We now explain the method. Consider repeating the single-qubit Ramsey cycle described in Sec.~\ref{sec:universal} with a fixed evolution time $\tau$.  The expectation value of a single-shot measurement at laboratory time $t$ is
\begin{equation}
P(t)=A\cos\left[2\pi\big(\nu_0+\delta\nu(t)\big)\tau\right]+B,
\end{equation}
where $\nu_0$ is an arbitrarily chosen value of the qubit energy (for example, approximate average offset with respect to the reference microwave) in units of $2\pi\hbar$ and $\delta\nu(t)$ is the qubit-energy fluctuation at laboratory time $t$. The constants $A$ and $B$ account for errors in the state preparation and measurement (SPAM).
Reference~\cite{Fink2013} does not consider errors, putting $A=1$ and $B=0$. We found that reflecting errors is crucial for the successful implementation of the method. 

The essence of the idea of Ref.~\cite{Fink2013} can be grasped from considering the time-correlator of single-shot measurements defined by\footnote{The function $C_P$ depends also on $\tau$, as is clear from Eq.~(3), but most of the time we write $C_P(t)$ instead of $C_P(t,\tau)$ for conciseness.}
\begin{equation}
\label{eq:CP}
C_P(t)\equiv\braket{P(t_0)P(t_0+t)}-\braket{P}^2,
\end{equation} 
where the angular brackets denote the time average (over $t_0$), equivalent to statistical ensemble average (assuming ergodicity), and we introduce the short hand notation $\braket{P} \equiv \braket{P(t_0)}$. 
In the quasi-static approximation and assuming the energy fluctuations are Gaussian \footnote{With Gaussian we mean noise whose statistical properties are fully determined by its mean and the two-point correlation function.}, the correlator is
\begin{subequations}
\begin{align}
\begin{aligned}
C_P(t)=\dfrac{A^2}{2}\bigg[&\cos(4\pi\bnu\tau)e^{-\frac{\chi_+(t)}{2}}+e^{-\frac{\chi_-(t)}{2}}\\&\ -2\cos^2(2\pi\bnu\tau) e^{-2\tau^2/T_2^{*2}}\bigg],
\end{aligned}
\label{eq:correlator_p}
\end{align}
where the coherence time and the envelope functions are
\begin{align}
\label{eq:T2*}
T_2^*&\equiv1/\pi\sqrt{2\braket{\delta\nu^2}},\\
\label{eq:envelope_func}
\chi_\pm(t)&\equiv8\pi^2\tau^2[\braket{\delta\nu^2}\pm\braket{\delta\nu(t')\delta\nu(t'+t)}].
\end{align}
\end{subequations}
Subtracting the squared mean $\braket{P}^2$ in the definition of $C_P(t)$ in Eq.~\eqref{eq:CP} serves to cancel out $B$ in Eq.~\eqref{eq:correlator_p}, so that the errors only enter through $A$ in the prefactor.

The next step is to note the different behavior of the two envelope functions. While they both go to a common constant at long times, $\chi_\pm(t\rightarrow\infty)=8\pi^2\tau^2\braket{\delta\nu^2}$, they differ at short times: $\chi_-(t\rightarrow0)=0$ in contrast to $\chi_+(t\rightarrow0)=16\pi^2\tau^2\braket{\delta\nu^2}$. For evolution time that satisfies $\tau\gg T_2^*/\sqrt{2}$, the first and third terms on the right-hand side of Eq.~\eqref{eq:correlator_p} can be dropped,
\begin{align}
C_P(t)\approx\frac{A^2}{2}e^{-\frac{\chi_-(t)}{2}}.
\label{eq:correlator_p_approx}
\end{align}
The authors of Ref.~\cite{Fink2013}  obtain this equation with $A=1$ and propose to do spectroscopy after extracting $\chi_-(t)$. We found that with SPAM errors it becomes difficult to reliably extract $\chi_-(t)$ from Eq.~\eqref{eq:correlator_p_approx}. We thus rewrite Eq.~\eqref{eq:correlator_p_approx} by taking logarithm and using the definition in Eq.~\eqref{eq:envelope_func},
\begin{align}
\braket{\delta\nu(t')\delta\nu(t'+t)}=\braket{\delta\nu^2}+\frac{1}{4\pi^2\tau^2}\left[\log C_P(t)-\log\frac{A^2}{2}\right],
\label{eq:correlator_nu}
\end{align}
The auto-PSD of the qubit energy noise, defined by
\begin{align}
S(f)=\int_{-\infty}^\infty dt\ \braket{\delta\nu(t')\delta\nu(t'+t)}e^{2\pi i f t},
\label{eq:psd}
\end{align}
then follows as
\begin{align}
S(f)=\frac{1}{4\pi^2\tau^2}\int_{-\infty}^\infty dt\ e^{2\pi i f t} \log C_P(t) + \mathrm{const} \times \delta(f).
\label{eq:psd-from-correlations}
\end{align}
The SPAM errors represented by $A$ only contribute to the irrelevant zero-frequency component, the last term. In sum, to obtain the noise PSD, we evaluate Eq.~\eqref{eq:psd-from-correlations} with the autocorrelation of single-shot measurements $C_P(t)$ estimated from data according to Eq.~\eqref{eq:CP}.

\begin{figure}
\centering
\includegraphics[width=\columnwidth]{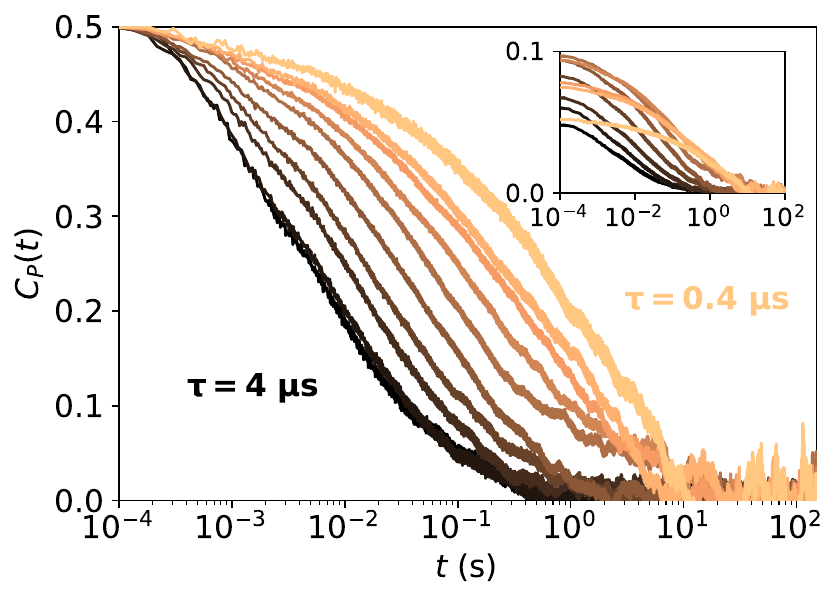}
\caption{Time-correlator of single-shot outcomes from the Ramsey experiment performed in qubit R of device D2. The color relates to the evolution time $\tau$ which changes from 0.4 $\mu$s to 4 $\mu$s in 0.4 $\mu$s steps. All curves are scaled to have a maximum at 0.5 to clearly show that correlations survive longer for shorter evolution times. If this scaling is not done each line has a different amplitude as is shown in the inset. }
\label{fig:correlator_p}
\end{figure}

We apply this method to qubit L of device D2, acquiring data for 5 minutes, with a cycle time of 43 $\mu$s for a single-shot measurement. We estimate the correlator $C_P(t)$ for several values of the fixed evolution time $\tau$ and plot the results in Fig.~\ref{fig:correlator_p}. While correlations decay with time as expected, we observe artifacts due to statistical noise at $t>0.1$ s. Since these artifacts affect the calculation of auto-PSD, we get rid of them by cropping the correlator, keeping only data for $t\leq0.1$ s. This value sets the lowest frequency accessible by the method, being here $f\geq5$ Hz. Due to the single-shot measurement, the correlator $C_P(t=0)$ is not estimated properly and needs to be corrected \cite{Malinowski2017}. For this we fit $C_P(t>0)$ with splines and replace $C_P(t=0)$ with the extrapolation from the fit. Despite that the traces $C_P(t)$ plotted in Fig.~\ref{fig:correlator_p} for different $\tau$ are different, they lead to the same auto-PSD, confirming Eq.~\eqref{eq:psd-from-correlations}. We demonstrate it in Appendix~\ref{app:verification}. As another cross-check of our assumptions, we plot a histogram of the qubit-energy fluctuations as an inset in Fig.~\ref{fig:wide} and observe that the qubit-energy fluctuations are well fit by a normal distribution. Even though it does not prove that the noise is Gaussian (in the sense of our definition), observing normal distribution is reassuring. 

To be able to use Eq.~\eqref{eq:psd-from-correlations}, the evolution time is limited both from below and from above. For the first, our derivations assumed $T_2^*/\sqrt{2} \ll \tau$. Since for this qubit and an integration time of 5 minutes, the average coherence time is 1.03 $\mu$s for, we only use data with $\tau\geq2\ \mu$s. With this, the terms neglected in Eq.~\eqref{eq:correlator_p_approx} are smaller than $10^{-3}$. On the other hand, for $\tau$ too large, $C_P$ becomes too small and is difficult to resolve reliably in the statistical noise. We restrict ourselves to $\tau\leq4\ \mu$s. There exists an unknown optimal $\tau_\mathrm{opt}$ that balances both the validity of the approximations as well as the visibility of $C_P$, but it is dependent on the specific noise being probed and thus cannot be determined beforehand. Therefore, even though it is not necessary to acquire data for different $\tau$ to implement the method, we advise to do so as it helps improve the final result despite ignoring $\tau_\mathrm{opt}$. For each evolution time we calculate the auto-PSD as the Fourier transform of Eq.~\eqref{eq:correlator_nu}, and we average the obtained auto-PSDs. The auto-PSD we extract has points equally spaced in the horizontal axis, which leads to an exponential increase in the density of points in the log-log scale of the typical PSD plot. To avoid this overcrowding of points and improve the presentation of the data in Fig.~\ref{fig:wide}, we divide the intermediate frequency range into 16 intervals equally spaced in the logarithmic scale axis and average all points that fall within each interval \footnote{No points fell in one of those 16 intervals and thus the displayed auto-PSD at intermediate frequencies consists of 15 points.}.

The result of the described procedure is shown as connected triangles in Fig.~\ref{fig:wide}. The auto-PSD obtained from correlations of single-shot measurements agrees well with Bayesian estimation and CPMG methods at frequencies where they overlap and interpolates the two consistently by a nontrivial shape: Going from lower to higher frequencies, the spectrum first continues to decay as $f^{-1.4}$, changes to a faster $f^{-2}$ decay around $f\sim1$ kHz, and finally flattens out to $f^{-0.4}$ somewhere between 10 and 100 kHz. 
We speculate that this complicated spectral shape is dominated by nuclear spins for most of the frequency range, except for the flat $f^{-0.4}$ high frequency range where it is likely due to charge noise, since a different decay $f^{-0.7}$ is observed in the other qubit (see Appendix~\ref{app:qubitR}). Verification of this conjecture would demand novel spectroscopy methods to access the cross-PSD at higher frequencies that we leave for future studies \cite{Szankowski2016}. More importantly, apart from just filling the gap, the consistency of the three methods highly increases their reliability and the plausibility of the obtained spectrum. We obtained a similarly consistent spectrum for qubit R in device {D2} (see Appendix~\ref{app:qubitR}). We conclude that  
by combining the three methods we could directly probe the qubit noise auto-PSD over 9 decades in frequency.

\section{Diffusion suppression from magnetic gradient}\label{sec:b_dependence}

Before concluding the article, we present the effects of the external magnetic field. While small, they are systematic, and thus we discuss them using data obtained on device D1. We extract the time dependence of the qubit energy for several values of the external magnetic field, $B=\{70,\ 230,\ 300,\ 370\}$ mT, by the Ramsey experiment described in Section~\ref{sec:universal}. At each value of the $B$ field, we have to retune the device for optimal PSB readout conditions, leading to different readout times and, thus, the cycle times. We opt to keep the number of points per record at 100 and adjust the number of records such that we probe the noise for approximately 8 hours in each case.

\begin{figure}
\centering
\includegraphics[width=0.95\columnwidth]{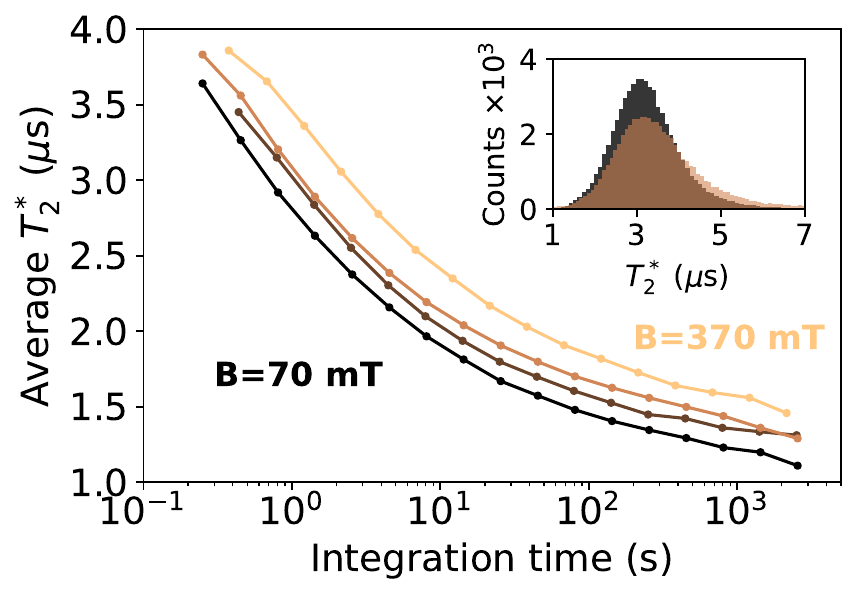}
\caption{Coherence time as a function of the integration times for qubit $R$ of Device D1. The four curves correspond to the magnetic field of 70 mT, 230 mT, 300 mT, and 370 mT, with a lower magnetic field displayed as a darker color. The corresponding interleaved Ramsey cycle times  $t_\mathrm{cycle}$ are $80\ \mu$s, $140\ \mu$s, $80\ \mu$s, and $120\ \mu$s, respectively. 
The inset shows sample histograms of coherence time values for two magnetic fields, {70 mT and 300 mT}, for an { integration time} of $0.25$ s.
}
\label{fig:B_dependence}
\end{figure}

Instead of plotting noise spectra on the logarithmic scale like in Fig.~\ref{fig:nuclear_psd} where small changes are difficult to spot, we examine  $T_2^*$ as the proxy\footnote{Ref.~\cite{Mehmandoost2024} says that the link is broken: $T_2^*$ is not given by the `total noise strength', if the latter is defined as the integral of PSD over all frequencies, equal to the variance.} to the overall noise strength. For the integration times that we reach, we are at the non-ergodic regime where $T_2^*$ is a stochastic quantity itself \cite{Delbecq2016}. The inset of Fig.~\ref{fig:B_dependence} illustrates this property, showing histograms of $T_2^*$.

We obtain them as follows. We split the data acquired over the $\sim8$ hours into blocks composed of $M$ records. Averaging data from one block yields a decaying oscillation that we fit to a curve $A+B\cos(2\pi\nu_0\tau)\exp[-(\tau/T_2^*)^2]$. We extract $T_2^*$ as a fitting parameter and average over all blocks for a given magnetic field. The average corresponds to the integration time equal to $M\times 100\times t_\mathrm{cycle}$. We work with logarithmically spaced values $M =\lfloor10^{1.5+0.25m}\rfloor$ varying $m=0,1,\ldots,m_\mathrm{max}$, choosing $m_\mathrm{max}$ such that we have at least {10} values of $T_2^*$ in each histogram.

The obtained average $T_2^*$ as a function of the integration time for different magnetic fields is displayed in the main panel of Fig.~\ref{fig:B_dependence}. The coherence time decreases with the integration time, and we expect it to continue to do so until an ergodic regime is reached. We did not reach that regime within our { total measurement time}. Nevertheless, a more interesting observation is a systematic improvement of the coherence time upon increasing the applied magnetic field. Going from 70 mT to 370 mT, the coherence time increases by about 30\%. We have seen similar behavior for qubit $L$ in the same device (not shown). The coherence increase is surprising and can not be due to a magnetic-field induced nuclear spin polarization: With the nuclear gyromagnetic factor for $^{29}$Si of $\gamma_n=-8.465$ MHz/T, the Zeeman splitting at $B=370$ mT is more than hundred times smaller than the thermal energy $k_B T$ at the dilution refrigerator operation temperature of  $~20$ mK.

We speculate that the increase might be due to the changes in the micromagnet field. We can determine its magnetic field $B^\mathrm{MM}_{L,R}$ acting on each qubit from the mean qubit energies, see Tab.~\ref{tab:mm_field}. We observe that the micromagnet is not fully polarized since upon increasing the external field the micromagnet field increases. With it, its difference at the two qubit sites also increases; see the last column of the table. We thus conclude that the magnetic field spatial gradient is probably also increasing. As is well known \cite{Walgraef1976,Genack1975}, a magnetic field gradient can suppress the nuclear spin diffusion by causing a mismatch of the Zeeman energies of a nuclear spin pair that would be otherwise free to undergo an energy-conserving flip-flop due to dipole-dipole interaction. The suppression of the nuclear spin diffusion has been observed in qubits defined in P donors in silicon, where the energy difference was due to the Knight field gradient \cite{Madzik2020}.

To get the magnetic-field spatial gradient, we would need to know the spatial separation of the qubits, and---to compare the gradients among different external fields---to know that the qubit-qubit distance does not change. Since we do not have reliable information on the qubits distance, we remain with the above qualitative observations. 

\begin{table}
\resizebox{\columnwidth}{!}{
\begin{tabular}{|c|c|c|c|c|c|c|}
\hline 
$\mathbf{B_\mathrm{ext}}$ \textbf{(mT)} & $\nu_L$ (MHz) & $\nu_R$ (MHz) & $B_L^\mathrm{MM}$ (mT) & $B_R^\mathrm{MM}$ (mT) & $\Delta B$ (mT) \\ 
\hline 
\textbf{70} & 4980 & 5121 & 107.8 & 112.8 & 5.0 \\ 
\hline 
\textbf{230} & 10521 & 10739 & 145.6 & 153.4 & 7.8 \\ 
\hline 
\textbf{300} & 12693 & 12921 & 153.1 & 161.3 & 8.2 \\ 
\hline 
\textbf{370} & 14808 & 15046 & 158.6 & 167.1 & 8.5 \\ 
\hline 
\end{tabular} }
\caption{Qubit energies and micromagnet field values for several external magnetic fields. 
}
\label{tab:mm_field}
\end{table}

\section{Conclusions}\label{sec:conc}

We have investigated noise affecting electron spin qubits in $^{\mathrm{nat}}$Si/SiGe quantum dots. We employed three different spectroscopy methods to directly probe the qubit energy noise spectrum for 9 orders of magnitude in frequency. One of the three methods, based on correlations of single-shot measurements, is novel and is crucial for accessing the mid-range frequencies, roughly between 10 Hz and 10 kHz.

 We found that the low-frequency ($f<10^2$ Hz) noise is similar across seven qubits in three different devices. This similarity---supported by additional theoretical estimates and experimental observations---suggests that the hyperfine coupling to the $^{29}$Si nuclear spins dominates here, with the spectrum falling of as a power-law $\propto f^{-1}$ that transforms into $\propto f^{-1.4}$ at around $f\approx0.3$ Hz. This power law differs from previously reported spectra on GaAs, which we attribute to the valley degree of freedom present in silicon. The spectrum shape then changes from $\propto f^{-1.4}$ into $\propto f^{-2}$ at about 1 kHz, and finally to a flatter $\propto f^{-\beta}$ with $\beta<1$ at about 10 kHz that we speculate is dominated by charge noise. 

Backed by the analysis of the cross-spectra, we identified device-dependent charge noise at low frequencies, which, while slightly weaker, is comparable to the dominant hyperfine noise. Finally, we observed an increase in the inhomogeneous dephasing time upon increasing the external magnetic field. We speculate that it is due to the suppression of the nuclear-spin diffusion by the magnetic gradient created by the not-fully-polarized micromagnet.

\acknowledgments
We thank for the financial support from CREST JST Grant No. JPMJCR1675 and from the Swiss National Science Foundation and NCCR SPIN Grant No. 51NF40-180604, JST Moonshot R\&D Grant Numbers JPMJMS226B and JPMJMS2065, JSPS KAKENHI Grant Nos. JP23H01790 and JP23H05455. J.S.R.-A. acknowledges support from the Gutaiteki Collaboration Seed, and K.T. support from JSPS KAKENHI grant No. 20H00237. We also acknowledge support from JST PRESTO Grant Numbers JPMJPR23F8 (A.N.), JPMJPR2017 (T.N.) and JPMJPR21BA (J.Y.).

\appendix

\section{Device specifics}\label{app:details}

In this section we give additional details on measurements and devices.
For D1, we use a low magnetic field, $B_\mathrm{ext}=70$ mT except for Sec.~\ref{sec:b_dependence} where it is 70, 230, 300, and 370 mT for the single-qubit Ramsey cycle time of 40, 70, 40, and 60 $\mu$s, respectively. The interleaved total Ramsey cycle time is the single-qubit cycle time multiplied by the number of qubits, being two in this device. The evolution time for the Ramsey sequence goes from $\tau_\mathrm{0}$ to $\tau_\mathrm{max}=4\ \mu$s in $\delta\tau=0.04\ \mu$s steps. The total measurement time is $\sim8$ hours for each value of magnetic field.

For D2 we use $B_\mathrm{ext}=545$ mT. The single-qubit Ramsey cycle is 43 $\mu$s, leading to $t_\mathrm{cycle}=86\ \mu$s total cycle time in this two-qubit device. The evolution time is taken the same as in device D1. The total measurement time is 1 hour.

Device D3 is operated at $B_\mathrm{ext}=570$ mT and has a much slower single-qubit Ramsey cycle time of {3.26888 ms} due to a different readout mechanism. Since this device includes three qubits we have {$t_{\mathrm{cycle}}=9.8064$ ms}. The Ramsey sequence sweeps the evolution time from $\tau_\mathrm{0}$ to $\tau_\mathrm{max}=2\ \mu$s in $\delta\tau=0.05\ \mu$s steps. The total measurement time is $\sim1.8$ hours.

\section{Charge sensor noise spectrum}\label{app:sensor}
\begin{figure}
\includegraphics[width=0.8\columnwidth]{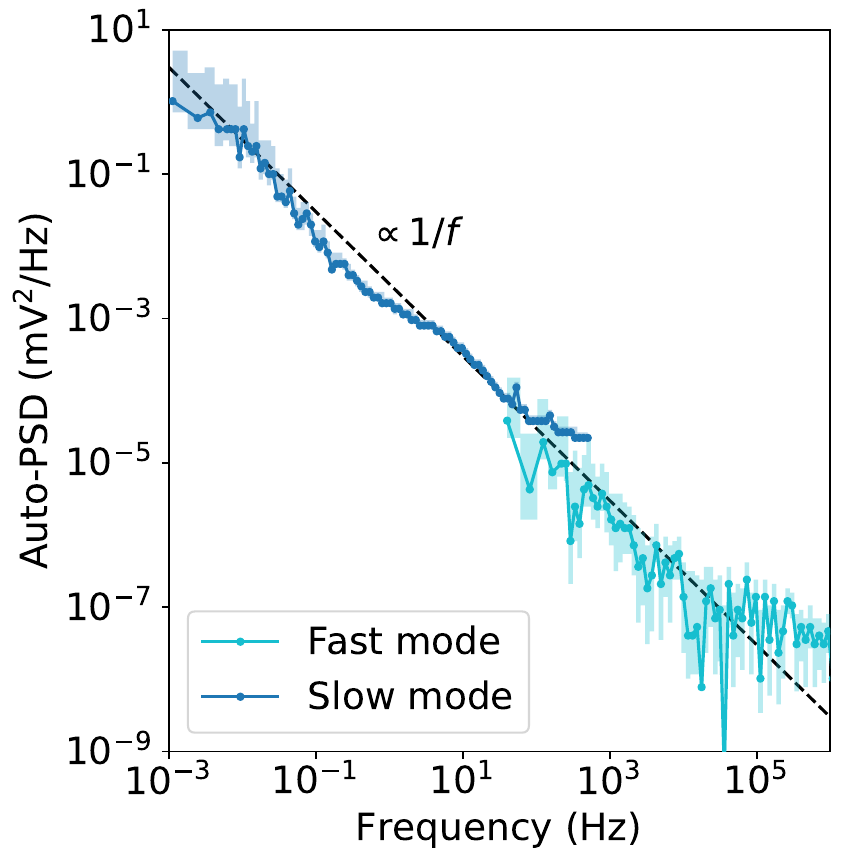}
\caption{Noise auto-spectrum for the charge sensor of device D1. The auto-PSD shown corresponds to the difference between the auto-PSDs of the charge sensor signal at a sensitive and insensitive conditions. The dashed line corresponds to a $f^{-1}$ dependence for reference. The points correspond to the most likely estimation of the PSD with the shaded regions denoting a 90\% confidence interval obtained by the methods in Ref.~\cite{Gutierrez-Rubio2022}.}
\label{fig:sensor_psd}
\end{figure}

As an additional check that spectra in Fig.~\ref{fig:nuclear_psd} are not dominated by charge noise, we additionally analyze the charge sensor signal in device D1. The signal corresponds to a voltage from a reflectometry circuit which we measure at both a sensitive (on one side of a Coulomb peak of the sensor QD) and insensitive (at a valley in between two Coulomb peaks of the sensor QD) condition. In this way, we are probing purely charge noise. In order to determine the noise spectrum over a broad frequency range, we use two measurement modes. First, a fast mode, where we acquire reflectometry voltages with a rate of 1 MHz for a time span of 0.1 s. Second, a slow mode, with an acquisition rate of 1 kHz for 1 hour. The charge noise auto-PSD is obtained through the difference of the auto-PSDs at the sensitive and insensitive conditions using the method explained in Appendix G of Ref.~\cite{Gutierrez-Rubio2022}. The resulting spectrum is shown in Fig.~\ref{fig:sensor_psd}. It is markedly different from spectra in Fig.~\ref{fig:nuclear_psd}. Given the similarity among the spectra of qubits in the same device, if the noise were charge dominated, we would expect the sensor noise to be also similar. Since it is not, the discrepancy provides further evidence that the qubit noise is dominated by nuclear spins.

\section{Nuclear spin diffusion model}\label{app:nuclear}

Assuming that the electron orbital ground state in the quantum dot confinement potential is a Gaussian, Refs.~\cite{Reilly2008,Malinowski2017,TaylorThesis} give the derivation of the noise auto-PSD originating from the hyperfine coupling to the nuclear spins. However, in silicon the wave function incorporates fast valley oscillations alongside the Gaussian envelope \cite{Hensen2020}:
\begin{equation}
|\psi(\vec{x})|^2\propto\exp\left(-\dfrac{x^2+z^2}{l_p^2}-\dfrac{y^2}{l_y^2}\right)\cos^2(k_vy),
\label{eq:wavefunction}
\end{equation}
where $l_p$ and $l_y$ are the in-plane and out-of-plane confinement lengths, and {$k_v=0.85/2\pi a_0$} with $a_0=0.543$ nm the lattice constant of Si. Repeating the derivations in Refs.~\cite{Reilly2008,Malinowski2017,TaylorThesis} with Eq.~\ref{eq:wavefunction} instead of a simple Gaussian ($k_v=0$), the time correlator of the qubit energy yields:
\begin{align}
\begin{split}
\braket{\nu(t')\nu(t'+t)}&=\frac{C_0\ (1+e^{-k_v^2l_y^2})^{-2}}{(1+\gamma|t|)(1+\gamma\zeta|t|)^{1/2}}\\
&\quad\times\bigg[1+2e^{-k_v^2l_y^2}e^{\frac{k_v^2l_y^2}{2(1+\gamma\zeta|t|)}}\\
&\qquad+\frac{e^{-2k_v^2l_y^2}}{2}\left(1+e^{\frac{2k_v^2l_y^2}{(1+\gamma\zeta|t|)}}\right)\bigg],
\end{split}
\label{eq:nuclear_correlator_valley}
\end{align}
where we defined $\gamma=2D/l_p^2$, $\zeta= l_p^2/l_y^2$, and:
\begin{equation}
C_0=p\dfrac{A^2}{h^2({2\pi})^{3/2}}\dfrac{I(I+1)}{3}\dfrac{a_0^3}{8(l_p^2l_y)},
\label{eq:C0}
\end{equation}
with $A=2.4\ \mu$eV the hyperfine coupling strength to the $^{29}$Si atoms \cite{Philippopoulos2020,Assali2011,Schliemann2003}, $I=1/2$ the nuclear spin, $D=1.6$ nm$^2$/s the diffusion constant \cite{Hayashi2008}, and $p=4.7\%$ the fraction of $^{29}$Si. The time correlator in the absence of the valley degree of freedom is obtained in the limit $k_v\rightarrow0$. Inserting Eq.~\ref{eq:nuclear_correlator_valley} into the definition in Eq.~\eqref{eq:psd} yields the black dashed line in Fig.~\ref{fig:nuclear_psd}.

If the harmonic in-plane confinement is parameterized by confinement energy $\hbar\omega_0$, the corresponding confinement length is $l_p\equiv\sqrt{\hbar/m\omega_0}$, with $m$ the effective electron mass in silicon. Using this in Eq.~\eqref{eq:C0} we find that noise power is proportional to the in-plane confinement energy and the square root of the out-of-plane confinement energy.

\section{Unnormalized cross-PSDs of device D3}\label{app:lorentzian}

\begin{figure}
\centering
\includegraphics[width=0.75\columnwidth]{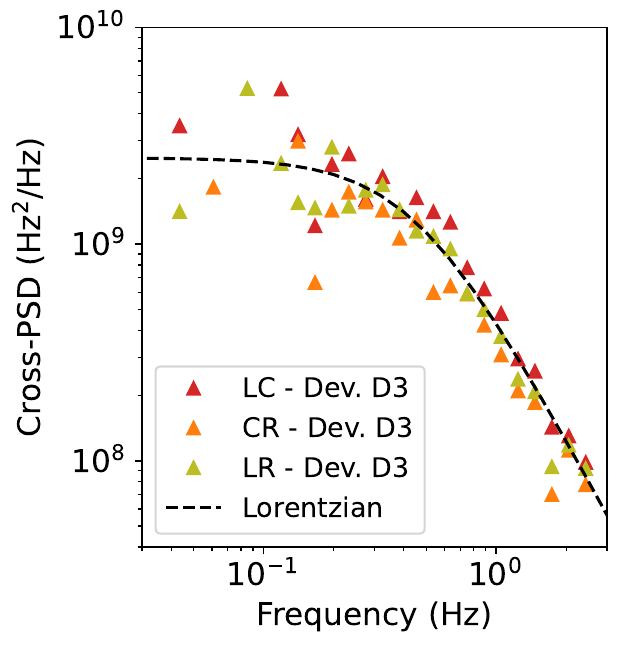}
\caption{Unnormalized magnitude of the three cross-PSDs measured in device D3. A Lorentzian curve is shown as a dashed line for reference.}
\label{fig:lor}
\end{figure}

To illustrate the TLS signature in the cross-PSDs of device D3, in Fig.~\ref{fig:lor} we show the unnormalized magnitude of the cross-PSDs from device D3. The correlations follow a Lorentzian shape $\propto1/(1+(2\pi f t_c)^2)$ which indicates coupling to a single TLS. A fit gives its switching time $t_c=0.35$ s. The cross-PSDs between first-neighbor qubits have the same amplitude as that of second neighbors, which we take as indication that the TLS responsible for these correlations is located near the qubits.

\section{Scaling of $C_P$ with $\tau$}\label{app:verification}

\begin{figure}
\centering
\includegraphics[width=\columnwidth]{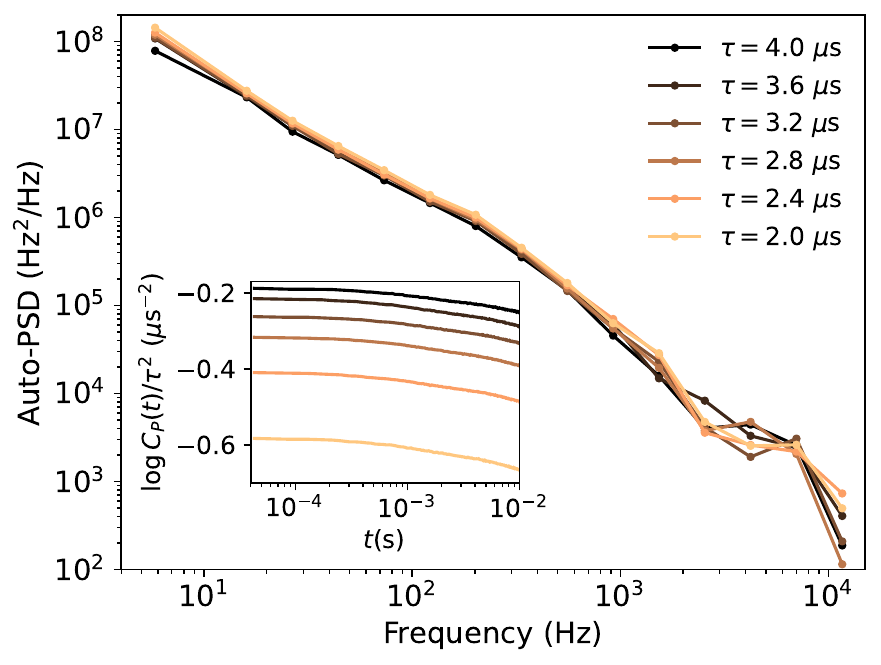}
\caption{Auto-PSDs from correlation of single-shot measurements obtained from data with different evolution times in qubit R of device D2. Inset: Logarithm of the correlator of single-shot readouts scaled by the evolution time $\tau$.}
\label{fig:logCp}
\end{figure}

When one looks at the traces $C_P(t)$ in Fig.~\ref{fig:correlator_p} for different $\tau$, it is not obvious that they yield the same power spectrum. The independence of the spectrum $S(f)$ on $\tau$ implied by Eq.~\eqref{eq:correlator_nu} requires the functional form $C_p(t,\tau) = g(\tau) h(t)^{\tau^2}$. To demonstrate it, in the inset of Fig.~\ref{fig:logCp} we plot $\log[C_P(t)]/\tau^2$ for the curves used in extracting the auto-PSD. Indeed, apart from an overall shift, the rescaled curves are very similar. Upon a Fourier transform the shift becomes an irrelevant zero-frequency component and we can extract the auto-PSDs from any individual curve, as shown in the main panel of Fig.~\ref{fig:logCp}. To benefit from the whole dataset, we use each curve and average the resulting Fourier transforms, obtaining the auto-PSDs plotted in Figs.~\ref{fig:wide} and \ref{fig:wide2}.

\section{Wide range noise spectroscopy of qubit R of device D2}\label{app:qubitR}

\begin{figure}
\centering
\includegraphics[width=\columnwidth]{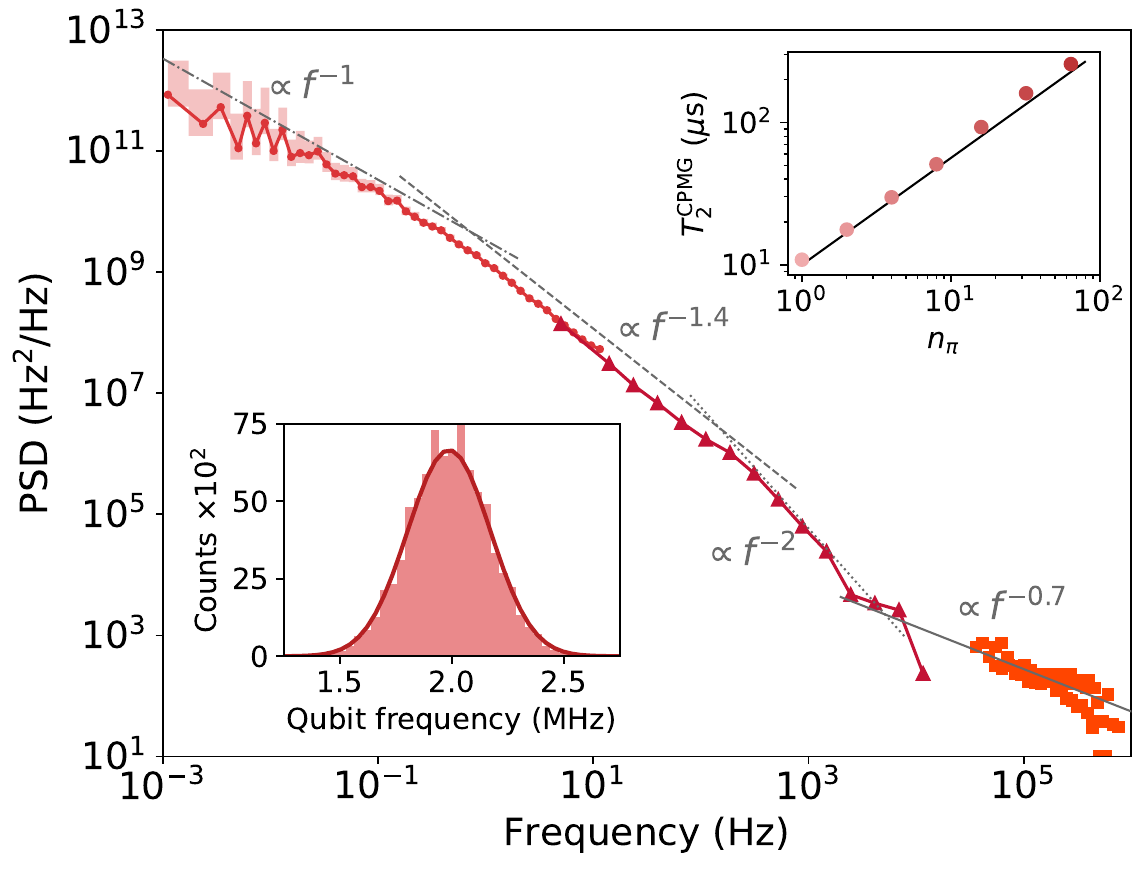}
\caption{Noise auto-PSD of qubit R of device D2. Symbols are analogous to those in Fig.~\ref{fig:wide} except for the solid gray line, which is here  proportional to $f^{-0.7}$. Upper right inset: coherence time for different number of pulses of the CPMG sequence. Lower left inset: histogram of qubit energies with a Gaussian fit as a continuous curve.}
\label{fig:wide2}
\end{figure}

To show the consistency of our noise spectroscopy methods, in Fig.~\ref{fig:wide2} we present the analog of Fig.~\ref{fig:wide} for qubit R of device D2. We observe a similar behavior to that of qubit L: The $f^{-1}$ power law transforms into $f^{-1.4}$, then becomes closer to $f^{-2}$ before becoming flatter at higher frequencies. In the insets we show the increasing coherence time with the number of dynamical decoupling pulses (upper right) as well as the Gaussian distribution of qubit frequencies in the histogram (lower left).

\clearpage
\bibliography{references.bib}

\end{document}